# Breaching the Human Firewall: Social engineering in Phishing and Spear-Phishing Emails

## Marcus Butavicius
National Security and Intelligence, Surveillance and Reconnaissance (ISR) Division
Defence Science and Technology Group
Edinburgh, South Australia
Email: marcus.butavicius@dsto.defence.gov.au

## Kathryn Parsons
National Security and Intelligence, Surveillance and Reconnaissance (ISR) Division
Defence Science and Technology Group
Edinburgh, South Australia
Email: kathryn.parsons@dsto.defence.gov.au

## Malcolm Pattinson
Business School
University of Adelaide
Adelaide, South Australia
Email: malcolm.pattinson@adelaide.edu.au

## Agata McCormac
National Security and Intelligence, Surveillance and Reconnaissance (ISR) Division
Defence Science and Technology Group
Edinburgh, South Australia
Email: agata.mccormac@dsto.defence.gov.au

## Abstract

We examined the influence of three social engineering strategies on users' judgments of how safe it is to click on a link in an email. The three strategies examined were *authority*, *scarcity* and *social proof*, and the emails were either genuine, phishing or spear-phishing. Of the three strategies, the use of authority was the most effective strategy in convincing users that a link in an email was safe. When detecting phishing and spear-phishing emails, users performed the worst when the emails used the authority principle and performed best when social proof was present. Overall, users struggled to distinguish between genuine and spear-phishing emails. Finally, users who were less impulsive in making decisions generally were less likely to judge a link as safe in the fraudulent emails. Implications for education and training are discussed.

## 1 Introduction

Phishing emails are emails sent with malicious intent that attempt to trick recipients into providing information or access to the sender. Typically, the sender masquerades as a legitimate entity and crafts the email to try and persuade the user to perform an action. This action may involve revealing sensitive personal information (e.g., passwords) and / or inadvertently providing access to their computer or network (e.g., through the installation of malware) (Aaron and Rasmussen 2010; APWG 2014; Hong 2012). In a recent survey of Australian organisations, the most common security incident reported (45%) was that of employees opening phishing emails (Telstra Corporation 2014). While the direct financial costs of such cyber-attacks in 2013 is estimated at a staggering USD $5.9 billion (RSA Security 2014), there are also a range of other negative consequences to organisations that can be just as harmful (Alavi et al. 2015). These include damage to reputation, loss of intellectual property and sensitive information, and the corruption of critical data (Telstra Corporation 2014).

A more sinister development in cyber-attacks has been the increase in spear-phishing (Hong 2012). In contrast with phishing emails, which tend to be more generic and are sent in bulk to a large number of recipients, spear-phishing emails are sent to, and created specifically for, an individual or small group of individuals (APWG 2014). When directed towards senior executives and high-ranking staff, such attacks are known as 'whaling'. These targets typically have greater access to sensitive corporate information and may have privileged access accounts when compared to the average user. Spear-phishing emails include more detailed contextual information to increase the likelihood of a recipient falling victim to them (Hong 2012). For example, they may include information relevant to the recipient's personal or business interests to increase the likelihood that the recipient will respond. Such attacks are increasingly deployed by criminals who are attempting to commit financial crimes against specific targets, corporate spies involved in stealing intellectual property and sensitive information, and hacktivists who wish to draw attention to their cause (APWG 2014).

Phishing and spear-phishing remain ongoing threats because they circumvent many technical safeguards by targeting the user, rather than the system (Hong 2012). Previous phishing studies have attempted to understand these human issues by studying the visual and structural elements of emails that influence people (Jakobbsson 2007; Furnell 2007; Parsons et al. 2013). However, phishing emails also frequently use social engineering to coerce the target into responding (Samani and McFarland 2015), and there is a lack of research examining the influence of social engineering strategies.

## 1.1 The Influence of Social Engineering Strategies

Social engineering refers to the psychological manipulation of people into disclosing information or performing an action (Mitnick et al 2002). This paper focuses on how three different social engineering strategies influence users' response to emails. To our knowledge, no previous studies on phishing have manipulated social engineering strategies in a controlled user study. However, psychological persuasion has been studied extensively in other contexts such as advertising and helping behaviour (Knowles and Lin 2004).

Although there is disagreement in the literature as to how to categorise persuasion strategies (Shadel and Park 2007; Pratkanis 2007), the most widely accepted classification of psychological persuasion strategies is by Cialdini (2007). Cialdini's (2007) summary includes six principles of persuasion. Three of the tactics, namely, reciprocation, consistency and liking, are more dependent on a mutual, recurring relationship, and are therefore less suited to the lab-based scenario of our study, in which the users do not have any relationship with the sender. Hence, our study focused on Cialdini's (2007) other three principles: *social proof*, *scarcity* and *authority*.

Social proof suggests that people are more likely to comply with a request if others have already complied (Cialdini 2007). In a phishing context, emails that specify the offer has already been taken up by other people are likely to be more persuasive. Scarcity is based on the idea that people are more likely to value something that is rare or limited. Hence, people are more likely to be influenced by emails that claim an offer is only available for a short time. The authority principle indicates that people are more likely to comply with a request that appears to be from a respected authority figure. Hence, an email with a request from the CEO of an organisation should be more effective than the same request from a less influential person. A recent survey of phishing emails reported that, between August 2013 and December 2013, authority was the most prevalent social engineering technique used in such attacks followed closely by scarcity, particularly in emails that requested information on account details (Akbar 2004).

## 1.2 The Design of Phishing Studies

There are only two previous studies on human susceptibility to phishing emails that have involved the direct manipulation of Cialdini's (2007) influence techniques. Neither of these were controlled user studies, and they yielded contradictory results. While Wright et al. (2014) found that authority was the least influential technique in tricking people into falling for phishing scams, Halevi et al. (2015) found the same strategy to be the most influential. Both these studies used the methodology that has been become known in the literature as 'real phishing', whereby users unknowingly receive emails as part of the test in their normal inbox (see also Jagatic et al. 2007).

To address these contradictory findings and further examine the issue of how Cialdini's techniques may influence people's susceptibility to phishing emails, we decided to use an alternative, lab-based approach (see also Parsons et al. 2013; Pattinson et al. 2012). Using this methodology, users were presented with emails in a controlled setting and their behaviour was logged. While such studies may lack the real-world face validity of 'real phishing' experiments, they have the advantage that they provide greater control and more comprehensive measurement of user behaviour. For example, 'real phishing' studies do not measure how people make decisions on genuine emails. In our study, by testing performance on both genuine and fraudulent emails, we can apply an approach to evaluation known as Signal Detection Theory (SDT: Green and Swets 1966). Previously, we applied this approach to the analysis of human detection performance of phishing emails (Parsons et al. 2013). SDT has also been applied to a wide range of other applications including human face recognition and image identification, biometric system assessment, economics and neuroscience (Butavicius 2006; Fletcher et al. 2008; Gold and Shadlen 2007; Hanton et al. 2010). SDT allows us to estimate two measures: *discrimination* and *bias*. Discrimination measures how well people can distinguish between genuine and fraudulent emails and bias measures people's tendency to classify an email as either genuine or fraudulent. In contrast, 'real phishing' studies cannot estimate discrimination and bias measures because they only collect behavioural responses to phishing emails[1].

## 1.3 Individual differences

Finally, we need to understand what makes some individuals better at detecting phishing emails than others. Halevi et al. (2015) and Pattinson et al. (2012) have shown that a number of individual differences (e.g., personality characteristics and familiarity with computers) can influence how people respond to phishing emails. In the current study, we included the Cognitive Reflection Test (CRT: Frederick 2005), which measures how impulsive people are when making decisions (i.e., cognitive impulsivity). Sagarin and Cialdini (2004) have argued that resisting persuasion techniques requires more cognitive resources than accepting them. Accordingly, previous research has shown that higher individual levels of impulsivity in decision making are associated with poorer performance in the detection of phishing emails (Parsons et al. 2013). In this study, we sought to replicate this finding. In addition, we tested whether cognitive impulsivity was associated with the ability to detect spear-phishing emails. By understanding what individual differences influence information security behaviours, we can begin to assist in the training and education of users to resist phishing attacks.

## 1.4 Aims of the research

In summary, the current study seeks to address the following questions:

- How do the three social engineering strategies of *authority*, *scarcity* and *social proof* influence users' judgments on the safety of links in emails?
- Does the influence of these techniques vary across different types of emails, i.e., *genuine*, *phishing* and *spear-phishing* emails?
- How well can people detect *phishing* and *spear-phishing* emails?
- How does a user's impulsivity in making decisions affect their ability to judge the safety of a link in an email?

In what follows, we will describe the methodology of our experiment, present a statistical analysis of the results of our study, and then discuss the findings and implications of this work, with a focus on training and education.

# 2 Methodology

## 2.1 Participants

Our convenience sample consisted of 121 students from a large South Australian university, and they were recruited via email invitation. At the time of the study, these students were currently enrolled in undergraduate and postgraduate level courses including finance, international business, accounting, marketing, management and entrepreneurship. Approximately half of the participants (60) had undertaken most of their tertiary education in Australia. The majority of the students were female (68%), all were 18 years or older and most were between 20-29 years of age (62%).

## 2.2 Emails

Our experiment used 12 emails, which were either genuine, phishing or spear-phishing emails. To create the emails, we consulted with university IT security staff, who provided examples of phishing emails that had been sent to university email accounts. These phishing emails had been sent within the previous six months and, based on recipient-reporting and system monitoring, had been identified as the most successful attacks against students and staff. These standard phishing emails (i.e., not spear-phishing) were used as templates to form the phishing emails in our experiment. For safety purposes, we disabled students' access to the internet during the study, and also modified the actual phishing link by a single character. We also collected genuine emails that had been received by students of the university to provide as a template for the remaining emails. These were used to create the genuine and spear-phishing emails in our study, where the only difference between the two emails was the link. For genuine emails we used a legitimate link, while for spear-phishing emails, we used the modified, illegitimate links from the actual phishing emails as previously specified.

---

[1] 'Real phishing' studies can only calculate *hits* (i.e., the number of correct detections of a phishing email) but not *false alarms* (i.e., the number of incorrect judgments of a genuine email as phishing).

In all phishing and spear-phishing emails, the displayed text for a link was a description such as "Click here" or "Take the survey" rather than the actual link, and participants were advised, both verbally and in writing at the start of experiment, that if they "hover over a link, it will show you where it would take you". Although the names and contact details in the emails were fictitious, the position titles in the genuine and spear-phishing emails were actual positions at the university. Participants were advised that, when judging the emails, they were to assume that the emails had all been sent to them deliberately (i.e., they had not received them by mistake) and that the topics in the emails were relevant to them (i.e., "if the email mentions a piece of software, assume that you are interested in that software").

In order to include the appropriate social engineering strategy, we added phrases to the emails that appealed to these strategies. There were four conditions testing the effects of social engineering:

- *Authority*: The email appeared to come from a person or institution of authority (e.g., CEO, CIO) and the language used was more authoritative.
- *Social proof*: The email encouraged the participants to take a particular action because other people, often peers, had already undertaken this action (e.g., "Over 1000 students will study overseas in 2014. Will you be one of them?").
- *Scarcity*: The email includes information that suggests an offer is limited (e.g., they have a limited time to respond or that there are a limited number of places available on a course).
- *None*: The email did not contain any phrases appealing to authority, social proof or scarcity strategies.

Each participant saw the same 12 emails. For each type of email (i.e., *genuine*, *phishing* and *spear-phishing*), we applied each of the four social engineering treatments once (i.e., *authority*, *social proof*, *scarcity* and *none*).

## 2.3 Procedure

Participants were allocated to separate lab-based sessions with a maximum of twenty students in each. Each student completed the experiment independently via computer. A research assistant explained the procedure and remained present in the room throughout the experiment to answer any questions and to ensure students worked independently. Participants were not explicitly told they were participating in an experiment on phishing. This is because previous research has shown that informing people they will view phishing emails artificially raises their awareness of phishing attacks for the duration of the experiment via a psychological process known as priming (Parsons et al. 2013). The experiment was delivered using the Qualtrics online survey software. Participants were shown each email separately and were asked to provide a 'Link Safety' judgment (i.e., 'It is okay to click on the link in this email'.). Responses were provided on a five point Likert scale where "1 = strongly disagree, "2" = disagree, "3" = neither agree nor disagree, "4" = agree and "5" = strongly agree. The emails were presented to participants in a different, random order for each session. After participants had judged all emails, they were then asked to complete the CRT in Qualtrics.

## 3 Results

To summarise the overall results, we recoded the 'Link Safety' judgments into a binary variable ('Safe to click?') such that scores of 4 ('agree') and 5 ('strongly agree') were classified as 'safe' and all remaining responses were classified as 'unsafe'. A summary of all participants' responses to links within the experiment can be seen in Figure 1. Participants correctly determined that legitimate links in genuine emails were safe to click 77% of the time. However, in spear-phishing emails, where the link was always unsafe, they incorrectly judged the link to be safe 71% of the time. Almost half the sample (45%) did not judge any of the links in the spear-phishing emails as unsafe. For standard phishing

emails, the percentage of responses that incorrectly judged the link to be safe dropped to 37%. 10% of the participants did not judge any of the links in the phishing emails as unsafe.

Next we analysed performance using a 4 x 3 Repeated Measures Analysis of Variance on the original five point 'Link Safety' ratings. There were four levels of *social engineering strategy*: *scarcity*, *social proof*, *authority* and *none*. There were three levels of *email type*: *genuine*, *phishing* and *spear-phishing*. As displayed in Figure 2, there was a significant overall influence of email type (Wilks' Lambda = .38, $F(2,119)$ = 80.09, $p < .001$, multivariate $\eta_p^2$ = .62). In other words, there was a significant variation in decisions on the safety of the link depending on whether the email was genuine, phishing or spear-phishing.

There was also a significant influence of social engineering strategy on the 'Link Safety' judgments of participants (Wilks' Lambda = .85, $F(3,118)$ = 6.89, $p < .001$, multivariate $\eta_p^2$ = .15). Overall, participants judged the links in phishing emails that conveyed authority as the safest (see Figure 2). Pairwise comparisons showed that the mean 'Link Safety' rating when the authority tactic was present was significantly higher than those for the other social engineering strategies (Mean $_{Authority - Scarcity}$ = .32, $CI_{95\%}$ = [.12, .525], SE = .08, $p < .001$; Mean $_{Authority - Social\ Proof}$ = .33, $CI_{95\%}$ = [.11, .55], SE = .08, $p < .05$). The mean for authority, although higher than the mean for emails with an absence of any social engineering strategy, was not significantly so (Mean $_{Authority - None}$ = .198, $CI_{95\%}$ = [-.02, .42], SE = .08, $p = .097$).

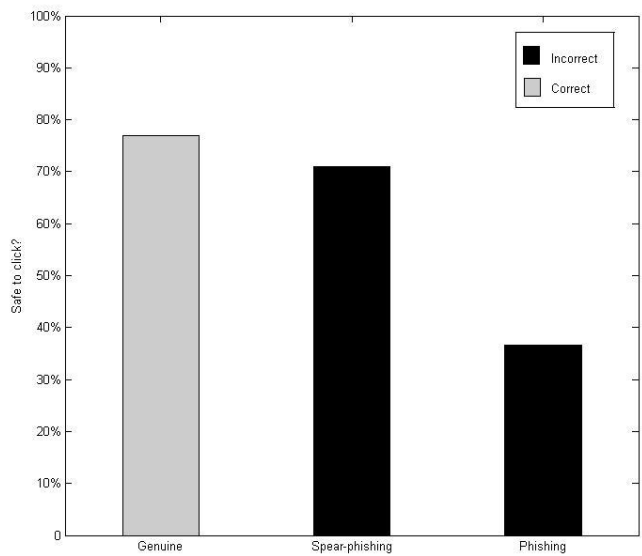

*Figure 1: Summary of 'Safe to click?' judgments across the experiment. Results are displayed for the three types of emails (genuine, spear-phishing and phishing). Correct responses are indicated in grey, while incorrect responses are shown in black.*

As can be seen in Figure 2, there was a significant interaction between the main effects of social engineering strategy and email type (Wilks' Lambda = .471, $F(6,115)$ = 21.5, $p < .001$, multivariate $\eta_p^2$ = .53). While the effect of social engineering strategy was similar for genuine and spear-phishing emails, the influence of those strategies was qualitatively different for phishing emails. Specifically, for phishing emails, when there was an absence of any social engineering strategy, the mean 'Link Safety' score was actually higher than when a social engineering strategy was present. In addition, the lowest mean rating was associated with the social proof attempts.

Using the Signal Detection Theory (SDT) approach, we calculated *A'* and *B"* which are non-parametric measures of *discrimination* and *bias*, respectively (Stanislaw & Todorov, 1999). *Discrimination* measures how well someone can distinguish between a fraudulent email and a genuine email. A score of 1 for *A'* means that discrimination ability is perfect while a score of 0.5 means that fraudulent emails cannot be distinguished from genuine emails. *Bias* measures someone's tendency to respond one way or the other, i.e., their bias towards saying an email is fraudulent or that it is genuine, regardless of how well they can discriminate between them. *B"* scores can range from -1 (everything is classified as fraudulent) to 1 (everything is classified as genuine) while zero indicates no response bias.

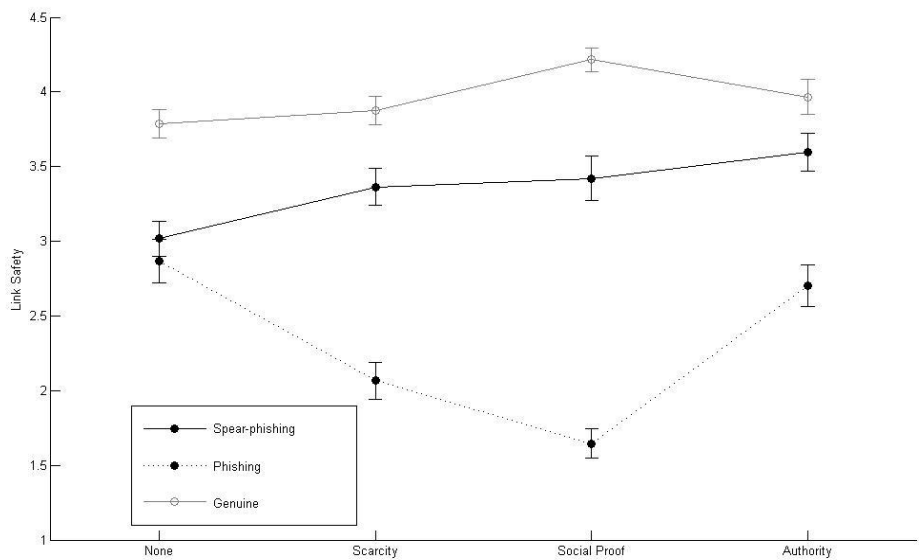

*Figure 2: Errorbar plot of means (+/- 1 SE) for 'Link Safety' ratings for each combination of Social Engineering Strategy (Y axis) and Deceit Effort (separate lines).*

First, we looked at these SDT measures in relation to detecting spear-phishing emails (see Table 1). According to the SDT framework, the decision-making task of the user is to distinguish between a 'signal' and 'noise' (Swets 1966). Using the binary variable 'Safe to click?', we defined 'noise' trials as cases where the email was genuine, and 'signal' trials as the spear-phishing emails. As described in Section 2.2, spear-phishing emails differed from the genuine emails only in the maliciousness of the embedded link. In this way, the 'signals' the user are trying to detect are the spear-phishing attempts, and the only cue is the legitimacy of the link. Secondly, we calculated SDT measures for users' ability to detect phishing emails. In this way, the 'signals' that the user are trying to detect are the phishing attacks and the distinguishing cues may include not just the hyperlink but also legitimacy of the sender, consistency and personalisation and spelling or grammatical irregularities (see Parsons et al. 2015).

|  |  | Authority | Scarcity | Social Proof | None |
|---|---|---|---|---|---|
| Spear-phishing | *A'* | 0.5 | 0.51 | 0.67 | 0.59 |
|  | *B"* | 0 | 0.01 | 0.25 | 0.07 |
| Phishing | *A'* | 0.72 | 0.82 | 0.90 | 0.67 |
|  | *B"* | 0.14 | 0.03 | 0.06 | 0.11 |

*Table 1. Signal Detection Theory measures for phishing and spear-phishing emails across the different social engineering conditions.*

Not surprisingly, users were better able to detect phishing emails (Mean *A'* = 0.78) than spear-phishing emails (Mean *A'* = 0.59). The relative effectiveness of the different social engineering strategies was the same for phishing and spear-phishing emails. Authority was the most successful strategy for confusing individuals as to the legitimacy of the fraudulent email and social proof was the least successful. When the fraudulent email used the authority strategy, participants were unable to reliably detect spear-phishing at all (*A'* = 0.5).

The major difference between the two analyses was for the emails with an absence of any social engineering strategy. For these emails, the ability to discriminate between genuine and fraudulent emails was relatively high for the spear-phishing emails (i.e., second best after social proof) whereas it was relatively lower for standard phishing emails (i.e., performance was worst of all conditions). In detecting phishing and spear-phishing attacks, users were biased towards responding that an email was legitimate in all but one of the conditions of the experiment.

Averaged 'Link Safety' judgments for individuals were compared against their scores on the CRT using Spearman's rank correlation coefficients ($\rho$). There was a significant negative correlation between CRT scores and link safety judgments for spear-phishing ($\rho = -.23$, *p* = .014, N = 112) and phishing ($\rho = -.3$, *p* = .001, N = 112) emails. In other words, participants who were less impulsive in decision-making were more likely to judge a link in a fraudulent email as unsafe. However there was no significant correlation between performance on the CRT and link safety judgments on genuine emails ($\rho = -.01$, *p* = .973, N = 114).

## 4   Discussion

In our study, the social engineering strategy that was most likely to influence users to judge that an email link was safe was authority, and the least effective strategy was social proof. The effectiveness of authority in our experiment, although in contrast with Wright et al.'s (2014) findings, supports the results of Halevi et al. (2015). In addition, our results concur with lab-based research into social engineering strategies in messages within emails (Guéguen and Jacob 2002) and marketing (Sagarin and Cialdini 2004) and are consistent with extensive research in other areas of psychology that suggest a strong tendency for people to be obedient towards authority (Blass 1999; Milgram 1974). The relative effectiveness of the social engineering strategies in our study was similar for both phishing and spear-phishing emails.

Overall, users demonstrated a bias towards classifying an email as genuine rather than fraudulent which is to be expected given that most emails in the wild are in fact genuine. Given the additional contextual information included in spear-phishing emails, it was also not surprising that participants were far worse at detecting them than the generic phishing emails in our experiment. However, what was alarming was the particularly poor performance of participants in trying to detect spear-phishing emails when appeals to authority were present. Taken as a whole, the participants were unable to reliably distinguish between spear-phishing and genuine emails when the email contained reference to an authority figure. What makes this particularly concerning is that:

a) the heightened effort and vigilance expected of users in a lab-based experiment should improve performance in comparison to real life,

b) participants were explicitly told how to check the real destination of a link in an email before the start of the experiment, and

c) the malicious link destinations were obviously unrelated to the content of the email.

Our findings are particularly worrying given the increase in spear-phishing in the wild reported in recent analyses of cyber-attacks (APWG 2014; Hong 2012; Samani and McFarland 2015) and the dominance of the authority persuasion technique within them (Akbar 2014). In fact, the success of such deceitful tactics, as demonstrated in our study, may partly explain their increased popularity.

Interestingly, the use of any social engineering technique in phishing emails appeared less effective than no technique at all. This may be due to an inoculation effect against this type of persuasion (McGuire 1970), whereby users have been exposed to so many generic phishing emails that attempt to use social engineering, that they have learnt to resist the persuasion attempt and not to respond to them. Such inoculation to persuasion has been demonstrated in marketing contexts (Friedstat and Wright 1994; Szybillo and Heslin 1973). However, a simpler explanation may also account for the findings. It may be that the presence of any social engineering strategy in standard phishing emails, where no significant effort has been made to target an individual using inside knowledge, is in fact a cue to the malicious intent behind the email. Without the necessary context, the persuasion may appear inappropriate and therefore raise the suspicions of the user.

Participants who were less impulsive in decision making were more likely to judge the links in phishing emails as more dangerous. Our findings replicated those of previous research that found that lower cognitive impulsivity was associated with resistance to phishing email attacks (Parsons et al. 2013). However, our results have also found that lower cognitive impulsivity can protect against targeted, high effort attacks such as spear-phishing. In addition, lower cognitive impulsivity did not adversely influence the judgments of genuine emails.

One of the limitations of our study was the use of a convenience sample of university students enrolled in subjects on business and information systems. Such a sample may not necessarily reflect the abilities of the wider population and, therefore, this limits the generalisability of our findings. As a result, we propose that future research should seek to replicate our study on a larger, more diverse sample.

The fact that our study found a potential link between someone's preference for a decision making style and their susceptibility to phishing has implications for future research in this area and, ultimately, the development of a training solution. Cognitive impulsivity is linked to what is known as dual processing models of persuasion (Chaiken et al. 1996). These models assume that we have two modes of processing information. The first mode, known as the 'central' mode, uses systematic processing that is highly analytical and detailed. The second mode, known as the 'peripheral' mode, is heuristic in nature and is more influenced by superficial cues. By using heuristics rather than detailed analysis, the 'peripheral mode' is faster and uses fewer cognitive resources than the 'central' mode. Humans have evolved to use a large number of heuristics that allow us to function effectively in a range of different scenarios (Gigerenzer et al. 1999). However, this efficiency comes at a cost because these heuristics are less-accurate than the analytical approach associated with the 'central' mode. The CRT measures someone's tendency to use the 'central' mode more than the 'peripheral' mode and therefore can account for the increase in errors in judging emails by people high in cognitive impulsivity.

Research has shown that our style of decision making can be modified, at least in the short term. For example, Pinillos et al. (2011) showed that completing the CRT itself can activate 'central' mode processing for subsequent tasks. Therefore, training people to defend against phishing attacks could focus on activating 'central' mode processing when people are judging emails. In the short-term, future research should investigate whether pre-testing with the CRT can improve phishing email discrimination. In the long-term, research may involve the development of structured analytic techniques similar in style to the techniques that are commonly used by intelligence analysts (e.g., Heuer 1999). Such techniques activate the 'central' mode of processing so that an analyst is less likely to make an incorrect assessment of intelligence by falling back on our natural tendency towards faster, but less accurate 'peripheral' processing. A possible training solution to phishing may require that we develop and teach analogous structured analytic techniques for assessing the legitimacy of emails. In other words, rather than simply warning users about the threat posed by malicious emails or providing them with specific examples, we may eventually be able to train people to use more effective strategies to detect phishing and spear-phishing attacks.

## Acknowledgements

This project was supported by a Premier's Research and Industry Fund granted by the South Australian Government Department of State Development.

## Copyright